\documentclass[12pt]{article}
\usepackage[dvips]{graphicx}
\usepackage{amsmath}
\usepackage{amsfonts}
\usepackage{amssymb}

\begin{document}

\author{S. Gov\thanks{Also with the Center for Technological Education Holon, 52
Golomb St., P.O.B 305, Holon 58102, Israel.} and S. Shtrikman\thanks{Also with
the Department of Physics, University of California, San Diego, La Jolla,
92093 CA, USA.}\\The Department of Electronics, \\Weizmann Institute of Science,\\Rehovot 76100, Israel
\and H. Thomas\\The Department of Physics and Astronomy,\\University of Basel,\\CH-4056 Basel, Switzerland}
\title{Magnetic trapping of neutral particles: Classical and Quantum-mechanical study
of a Ioffe-Pritchard type trap.}
\date{}
\maketitle
\begin{abstract}
Recently, we developed a method for calculating the \emph{lifetime} of a
particle inside a magnetic trap with respect to spin flips, as a first step in
our efforts to understand the quantum-mechanics of magnetic traps. The 1D toy
model that was used in this study was physically \emph{unrealistic} because
the magnetic field was \emph{not} curl-free. Here, we study, both classically
and quantum-mechanically, the problem of a neutral particle with spin $S$,
mass $m$ and magnetic moment $\mu$, moving in 3D in an inhomogeneous magnetic
field corresponding to traps of the Ioffe-Pritchard, `clover-leaf' and
`baseball' type. Defining by $\omega_{p}$, $\omega_{z}$ and $\omega_{r}$ the
precessional, the axial and the lateral vibrational frequencies, respectively,
of the particle in the adiabatic potential $V_{eff}=\mu\left|  \mathbf{B}%
\right|  $, we find classically the region in the $\left(  \omega_{r}%
/\omega_{p}\right)  $-$\left(  \omega_{z}/\omega_{p}\right)  $ plane where the
particle is trapped.

Quantum-mechanically, we study the problem of a spin-one particle in the same
field. Treating $\omega_{r}/\omega_{p}$ and $\omega_{z}/\omega_{p}$ as small
parameters for the perturbation from the adiabatic Hamiltonian, we derive a
closed-form expression for the transition rate $1/T_{esc}$ of the particle
from its trapped ground-state. In the extreme cases the expression for
$1/T_{esc}$ reduces to
\[
\dfrac{1}{T_{esc}}\simeq\left\{
\begin{array}
[c]{c}%
4\pi\omega_{r}\exp\left[  -\dfrac{2\omega_{p}}{\omega_{r}}\right]  \text{; for
}\omega_{p}\gg\omega_{r}\gg\omega_{z}\\
8\sqrt{2\pi}\sqrt{\omega_{p}\omega_{i}}\exp\left[  -\dfrac{2\omega_{p}}%
{\omega_{i}}\right]  \text{ ; for }\omega_{p}\gg\omega_{r}=\omega_{z}%
\equiv\omega_{i}\\
\sqrt{\dfrac{\pi}{2}}\omega_{r}\left(  \dfrac{\omega_{z}}{\omega_{p}}\right)
^{3/2}\exp\left[  -\dfrac{2\omega_{p}}{\omega_{z}}\right]  \text{; for }%
\omega_{p}\gg\omega_{z}\gg\omega_{r}%
\end{array}
\right.  .
\]
\end{abstract}

\section{Introduction.\label{intro}}

\subsection{Magnetic traps for neutral particles.\label{traps}}

Recently there has been rapid progress in techniques for trapping samples of
neutral atoms at elevated densities and extremely low temperatures. The
development of magnetic and optical traps for atoms has proceeded in parallel
in recent years, in order to attain higher densities and lower temperatures
\cite{t1,t2,t3,t4,t6}. We should note here that traps for neutral particles
have been around much longer than their realizations for neutral atoms might
suggest, and the seminal papers for neutral particles trapping as applied to
neutrons and plasmas date from the sixties and seventies. Many of these papers
are referenced by the authors of Refs.\cite{t1,t2,t3}. In this paper we
concentrate on the study of \emph{magnetic} traps. Such traps exploit the
interaction of the magnetic moment of the atom with the inhomogeneous magnetic
field to provide spatial confinement.

Microscopic particles are not the only candidates for magnetic traps. In fact,
a vivid demonstration of trapping large scale objects is the hovering magnetic
top \cite{levitron,ucas,harrigan,patent}. This ingenious magnetic device,
which hovers in mid-air for about 2 minutes, has been studied in the past few
years by several authors \cite{edge,Berry,bounds,simon,dynamic,dynamic2}.

\subsection{Qualitative description.\label{desc}}

The physical mechanism underlying the operation of magnetic traps is the
adiabatic principle. The common way to describe their operation is in terms of
\emph{classical} mechanics: As the particle is released into the trap, its
magnetic moment points antiparallel to the direction of the magnetic field.
Inside the trap, the particle experiences translation oscillations with
vibrational frequencies $\omega_{vib}$ which are small compared to its
precession frequency $\omega_{prec}$. Under this condition the spin of the
particle may be considered as experiencing a \emph{slowly} rotating magnetic
field. Thus, the spin precesses around the \emph{local} direction of the
magnetic field $\mathbf{B}$ (adiabatic approximation) and, on the average, its
magnetic moment $\mathbf{\mu}$ points \emph{antiparallel} to the local
magnetic field lines. Hence, the magnetic energy, which is normally given by
$-\mathbf{\mu}\cdot\mathbf{B}$, is now given (for small precession angle) by
$\mu\left|  \mathbf{B}\right|  $. Thus, the overall effective potential seen
by the particle is
\begin{equation}
V_{eff}\simeq\mu\left|  \mathbf{B}\right|  . \label{energy}%
\end{equation}
In the adiabatic approximation, the spin degree of freedom is rigidly coupled
to the translational degrees of freedom, and is already incorporated in
Eq.(\ref{energy}) such that the particle may be considered as having only
translational degrees of freedom. When the strength of the magnetic field
possesses a\emph{\ minimum}, the effective potential becomes attractive near
that minimum, and the whole apparatus acts as a trap.

As mentioned above, the adiabatic approximation holds whenever $\omega
_{prec}\gg\omega_{vib}$. As $\omega_{prec}$ is inversely proportional to the
spin, this inequality can be satisfied provided that the spin of the particle
is sufficiently small. If, on the other hand, the spin of the particle is too
large, it cannot respond fast enough to the changes of the direction of the
magnetic field. In this limit $\omega_{prec}\ll\omega_{vib}$, the spin has to
be considered as fixed in space and, according to Earnshaw's theorem
\cite{earnshaw}, becomes unstable against \emph{translations}. Note also that
$\omega_{prec}$ is proportional to the field $\left|  \mathbf{B}\right|  $. To
prevent $\omega_{prec}$ of becoming too small, resulting in spin-flips
(Majorana transitions), most magnetostatic traps include a \emph{bias} field,
so that the effective potential $V_{eff}$ possesses a \emph{nonvanishing} minimum.

\subsection{The purpose and structure of this paper.\label{purp}}

The discussion of magnetic traps in the literature is, almost entirely, done
in terms of \emph{classical} mechanics. In microscopic systems, however,
quantum effects become important, giving rise to a finite lifetime of the
particle within the trap. This requires a quantum-mechanical treatment
\cite{quant}. An even more interesting issue is the understanding of how the
classical and the quantum descriptions of a \emph{given} system are related.
It is important to note here that there are \emph{two} mechanisms by which the
particle can escape from the trap: The first one is the usual tunneling of the
particle from the trap, without a change of its spin state, to regions where
the magnetic field decreases to zero. The time scale for this process can be
evaluated by standard methods. The second way the particle can escape from the
trap is by flipping its spin state (Majorana transitions). This process, which
is different from the first one because there is no potential barrier, is the
subject of this paper.

As a first step in our efforts to understand the quantum-mechanics of magnetic
traps, we recently developed a method for calculating the \emph{lifetime} of a
particle inside a magnetic trap with respect to such a spin flip process
\cite{life1d}. The toy model that was used in this study consisted of a
particle with spin, having only a single translational degree of freedom, in
the presence of a 1D inhomogeneous magnetic field. We found that the trapped
state of the particle decays with a lifetime given by $\sim1/\left(  \sqrt
{K}\omega_{vib}\right)  \exp\left(  2/K\right)  $ where $K=\omega_{vib}%
/\omega_{prec}$, and where the result is valid for $K\ll1$. Though the field
that was used in this model did trap the particle, it was not \emph{realistic}
in the sense that it was not curl-free. Our next step was to study the case of
a particle with spin, having \emph{two} translational degree of freedom, in
the presence of a physically more realistic trapping field that, in
contradistinction to the toy model, \emph{is} curl-free \ \cite{life2d}. This
model is reminiscent of a Ioffe-Pritchard trap \cite{prit,t2,ife}, but without
the axial translational degree of freedom. Here we found that the lifetime is
given by $\sim1/\omega_{vib}\exp\left(  2/K\right)  $ which is similar to the
result found in the 1D case. In the present paper we describe an analysis of a
Ioffe-Pritchard type trap which \emph{includes} the axial translational degree
of freedom. We neglect the effect of interactions between the particles in the
trap, and we analyze the dynamics of a \emph{single} particle inside the trap.
Unlike in our previous papers, where we studied the case of a spin $1/2$
particle, we treat here the case of a spin $1$ particle, both as an example to
show the validity of our approach for higher spins, and also because it is
more relevant in view of the recent development in Bose-Einstein condensation experiments.

The structure of this paper is as follows: In Sec.(\ref{def}) we start by
defining the system we study, together with useful parameters that will be
used throughout this paper. Next, we carry out a \emph{classical} analysis of
the problem in Sec.(\ref{class}). Here, we find two stationary solutions for
the particle inside the trap. One of them corresponds to a state whose spin is
\emph{parallel} to the direction of the magnetic field whereas the other one
corresponds to a state whose spin is \emph{antiparallel }to that direction.
When considering the dynamical stability of these solutions, we find that only
the \emph{antiparallel }stationary solution is stable, as expected from the
discussion in Sec.(\ref{desc}) above. In Sec.(\ref{quant}) we reconsider the
problem from a \emph{quantum-mechanical} point of view for a spin-\emph{one}
particle. Here, we find states that refer to \emph{antiparallel} ($M=-1$,
where $M$ is the magnetic quantum number) and \emph{orthogonal} ($M=0$)
orientations of the spin, the first of these being bounded while the second
one is unbounded. We argue that the third possible situation, in which the
spin is \emph{parallel} ($M=+1$) to the direction of the field, has negligible
coupling to the bound state, and therefore can be neglected. We show that the
\emph{antiparallel} and \emph{orthogonal} states are \emph{coupled} due to the
inhomogeneity of the field, and we calculate the transition rate from the
bound state to the unbounded state. Finally, in Sec.(\ref{dis}) we compare the
results of the classical analysis with those of the quantum analysis and
comment on their implications for practical magnetic traps.

\section{Description of the problem.\label{def}}

We consider a particle of mass $m$, magnetic moment $\mu$ and intrinsic spin
$S$ (aligned with $\mu$) moving in an inhomogeneous magnetic field
$\mathbf{B}$ corresponding to traps of the Ioffe-Pritchard, `clover-leaf' and
`baseball' type \cite{t2}, and given by
\begin{align}
\mathbf{B}  &  =\left[  B_{0}+\dfrac{1}{2}B^{\prime\prime}z^{2}-\dfrac{1}%
{4}B^{\prime\prime}\left(  x^{2}+y^{2}\right)  \right]  \mathbf{\hat{z}%
}\label{d0}\\
&  +\left(  B^{\prime}-\dfrac{1}{2}B^{\prime\prime}z\right)  x\mathbf{\hat
{x}+}\left(  -B^{\prime}-\dfrac{1}{2}B^{\prime\prime}z\right)  y\mathbf{\hat
{y}}\text{.}\nonumber
\end{align}
This field possesses a nonzero minimum of amplitude at the origin, which is
the essential part of the trap. The Hamiltonian for this system is%
\begin{equation}
H=\dfrac{\mathbf{P}^{2}}{2m}-\mathbf{\mu\cdot B} \label{d0.1}%
\end{equation}
where $\mathbf{P}$ is the momentum of the particle.

We define $\omega_{p}$ as the precessional frequency of the particle when it
is at the origin $(x=0,y=0,z=0)$. Since at that point the magnetic field is
$\mathbf{B=}B_{0}\mathbf{\hat{z}}$ we find that
\begin{equation}
\omega_{p}\equiv\dfrac{\mu B_{0}}{S}\text{.} \label{d1}%
\end{equation}
Next, we define $\omega_{z}$ and $\omega_{r}$ as the small-amplitude axial and
lateral vibrational frequencies of the particle when it is placed with
\emph{antiparallel }spin into the adiabatic potential given by
\[
V_{eff}=\mu\left|  \mathbf{B}\right|  \simeq\mu B_{0}\left(  1+\dfrac
{B_{0}B^{\prime\prime}}{2B_{0}^{2}}z^{2}+\left(  \dfrac{(B^{\prime})^{2}%
-\frac{1}{2}B_{0}B^{\prime\prime}}{2B_{0}^{2}}\right)  r^{2}\right)
+\mathcal{O}\left(  x^{4},x^{2}y^{2},y^{4}\right)  ,
\]
from which we get
\begin{align}
\omega_{z}  &  \equiv\sqrt{\dfrac{\mu B^{\prime\prime}}{m}}\label{d2}\\
\omega_{r}  &  \equiv\sqrt{\dfrac{\mu\left[  (B^{\prime})^{2}-\frac{1}{2}%
B_{0}B^{\prime\prime}\right]  }{mB_{0}}}\text{.}\nonumber
\end{align}
In what follows we assume that $(B^{\prime})^{2}-\frac{1}{2}B_{0}%
B^{\prime\prime}>0$ such that $\omega_{r}$ is real. We also define the
ratios,
\begin{align}
K_{z}  &  \equiv\dfrac{\omega_{z}}{\omega_{p}}=\sqrt{\dfrac{B^{\prime\prime
}S^{2}}{\mu mB_{0}^{2}}}\label{d3}\\
K_{r}  &  \equiv\dfrac{\omega_{r}}{\omega_{p}}=\sqrt{\dfrac{\left[
(B^{\prime})^{2}-\frac{1}{2}B_{0}B^{\prime\prime}\right]  S^{2}}{\mu
mB_{0}^{3}}}\nonumber
\end{align}
These parameters will be our `measure of adiabaticity'. It is clear that as
$K_{z}$ and $K_{r}$ become smaller and smaller, the adiabatic approximation
becomes more and more accurate. Note that when the bias field $B_{0}$
vanishes, both $K_{z}$ and $K_{r}$ become infinite, and the adiabatic
approximation fails. We will show below that, under this condition, the system
becomes \emph{unstable }against spin flips, which is in agreement with our
discussion at the beginning. This shows that the introduction of the bias
field $B_{0}$, is \emph{essential }to the operation of the trap with regard to spin-flips.

\section{Classical analysis.\label{class}}

\subsection{The stationary solutions.\label{stat}}

We denote by $\mathbf{\hat{n}}$ a unit vector in the direction of the spin
(and the magnetic moment). Thus, the equations of motion for the center of
mass of the particle are
\begin{align}
m\dfrac{d^{2}x}{dt^{2}}  &  =\mu\dfrac{\partial}{\partial x}\left(
\mathbf{\hat{n}\cdot B}\right) \label{c1}\\
m\dfrac{d^{2}y}{dt^{2}}  &  =\mu\dfrac{\partial}{\partial y}\left(
\mathbf{\hat{n}\cdot B}\right) \nonumber\\
m\dfrac{d^{2}z}{dt^{2}}  &  =\mu\dfrac{\partial}{\partial z}\left(
\mathbf{\hat{n}\cdot B}\right) \nonumber
\end{align}
and the evolution of its spin is determined by
\begin{equation}
S\dfrac{d\mathbf{\hat{n}}}{dt}=\mu\mathbf{\hat{n}\times B}\text{.} \label{c2}%
\end{equation}
The two equilibrium solutions to Eqs.(\ref{c1}) and (\ref{c2}) are
\begin{equation}
\mathbf{\hat{n}}(t)=\mp\mathbf{\hat{z}} \label{c3}%
\end{equation}
with%
\begin{align*}
x(t)  &  =0\\
y(t)  &  =0\\
z\left(  t\right)   &  =0
\end{align*}
representing a motionless particle at the origin with its magnetic moment (and
spin) pointing \emph{antiparallel} ($\mathbf{\hat{n}}(t)=-\mathbf{\hat{z}}$)
to the direction of the field at that point and a similar solution but with
the magnetic moment pointing \emph{parallel} to the direction of the field
($\mathbf{\hat{n}}(t)=+\mathbf{\hat{z}}$).

\subsection{Stability of the solutions.}

To check the stability of these solutions we now add first-order
perturbations. We set
\begin{align}
\mathbf{\hat{n}(}t\mathbf{)}  &  =\mathbf{\mp\hat{z}+}\epsilon_{x}%
(t)\mathbf{\hat{x}+}\epsilon_{y}(t)\mathbf{\hat{y}}\label{c4}\\
x(t)  &  =0+\delta x(t)\nonumber\\
y(t)  &  =0+\delta y(t)\nonumber\\
z(t)  &  =0+\delta z(t)\nonumber
\end{align}
(note that, to first order, the perturbation $\delta\mathbf{\hat{n}}%
=\epsilon_{x}(t)\mathbf{\hat{x}+}\epsilon_{y}(t)\mathbf{\hat{y}}$ is
\emph{orthogonal} to the vector $\mathbf{\hat{n}}$ for the stationary solution
$\mathbf{\hat{n}}_{0}=\mathbf{\mp\hat{z}}$, since $\mathbf{\hat{n}}$ is a unit
vector), substitute these into Eqs.(\ref{c1}) and (\ref{c2}), and retain only
first-order terms. We find that the resulting equations for $\delta x(t)$,
$\delta y(t)$, $\delta z\left(  t\right)  $, $\epsilon_{x}(t)$ and
$\epsilon_{y}(t)$ are
\begin{align}
m\dfrac{d^{2}\delta x}{dt^{2}}  &  =\pm\dfrac{1}{2}\mu B^{\prime\prime}\delta
x+\mu B^{\prime}\epsilon_{x}\label{c5}\\
m\dfrac{d^{2}\delta y}{dt^{2}}  &  =\pm\dfrac{1}{2}\mu B^{\prime\prime}\delta
y-\mu B^{\prime}\epsilon_{y}\nonumber\\
m\dfrac{d^{2}\delta z}{dt^{2}}  &  =\mp\mu B^{\prime\prime}\delta z\nonumber\\
S\dfrac{d\epsilon_{x}}{dt}  &  =\mu B_{0}\epsilon_{y}\mp\mu B^{\prime}\delta
y\nonumber\\
S\dfrac{d\epsilon_{y}}{dt}  &  =-\mu B_{0}\epsilon_{x}\mp\mu B^{\prime}\delta
x.\nonumber
\end{align}
The motion of the $z$-coordinate is decoupled from the others. If
$B^{\prime\prime}>0$, it is stable only when the upper sign is taken,
corresponding to a spin \emph{antiparallel }to the direction of the field. It
can be shown that when $B^{\prime\prime}<0$ then, even if the system is stable
under axial vibrations (by choosing the lower sign), it cannot be stable as a
whole. We therefore disregard the lower sign, and the equation for the
$z$-coordinate for the rest of the derivation.

The normal modes of the reduced system transform as the irreducible
representations of the symmetry group. The 4-dimensional linear space spanned
by the deviations $(\delta x,\delta y,\epsilon_{x},\epsilon_{y})$ from the
stationary state carries the irreducible representations $\Gamma_{+}$ with
characters $e^{-i\gamma}$ and $\Gamma_{-}$ with characters $e^{+i\gamma}$, and
may thus be decomposed into the two 2-dimensional invariant subspaces
transforming as $\Gamma_{+}$ and $\Gamma_{-}$, respectively. These subspaces
are spanned by the circular position coordinates and precessional spin
coordinates%
\begin{align}
\Gamma_{+}:\;  &  (\rho_{+}=\delta x+i\delta y,\epsilon_{-}=\epsilon
_{x}-i\epsilon_{y});\label{c5.1}\\
\Gamma_{-}:\;  &  (\rho_{-}=\delta x-i\delta y,\epsilon_{+}=\epsilon
_{x}+i\epsilon_{y}). \label{c5.2}%
\end{align}
Thus, the normal modes consist of a circular motion in the $(x,y)$-plane
coupled to a precession of the spin vector in the \emph{opposite} sense.

Indeed, after introducing the $(\rho_{\pm},\epsilon_{\mp})$-coordinates into
Eqs.(\ref{c5}), this set of four equations decomposes into one pair of
equations for $(\rho_{+},\epsilon_{-})$ and another pair for $(\rho
_{-},\epsilon_{+})$. We now look for oscillatory (stable) solutions of these
equations and set
\begin{equation}
\rho_{\pm}=\rho_{\pm,0}e^{-i\omega t},\quad\epsilon_{\pm}=\epsilon_{\pm
,0}e^{-i\omega t}. \label{rhoe}%
\end{equation}
This yields the algebraic equations
\begin{equation}
\Gamma_{+}:\;\left(
\begin{array}
[c]{cc}%
\dfrac{1}{2}\mu B^{\prime\prime}+m\omega^{2} & \mu B^{\prime}\\
i\mu B^{\prime} & i\left(  \omega S+\mu B_{0}\right)
\end{array}
\right)  \cdot\left(
\begin{array}
[c]{l}%
\rho_{+,0}\\
\epsilon_{-,0}%
\end{array}
\right)  =\left(
\begin{array}
[c]{l}%
0\\
0
\end{array}
\right)  , \label{c8.1}%
\end{equation}
$\allowbreak$%
\begin{equation}
\Gamma_{-}:\;\left(
\begin{array}
[c]{cc}%
\dfrac{1}{2}\mu B^{\prime\prime}+m\omega^{2} & \mu B^{\prime}\\
-i\mu B^{\prime} & i\left(  \omega S-\mu B_{0}\right)
\end{array}
\right)  \cdot\left(
\begin{array}
[c]{l}%
\rho_{-,0}\\
\epsilon_{+,0}%
\end{array}
\right)  =\left(
\begin{array}
[c]{l}%
0\\
0
\end{array}
\right)  . \label{c8.2}%
\end{equation}
These equations have non-trivial solutions whenever the determinant of either
of the two matrices vanishes. This yields the secular equations%

\begin{align}
\Gamma_{+}  &  :\left(  \frac{\omega}{\omega_{p}}\right)  ^{3}+\left(
\frac{\omega}{\omega_{p}}\right)  ^{2}+\dfrac{1}{2}K_{z}^{2}\left(
\frac{\omega}{\omega_{p}}\right)  -K_{r}^{2}=0,\label{c11.1}\\
\Gamma_{-}  &  :\;\left(  \frac{\omega}{\omega_{p}}\right)  ^{3}-\left(
\frac{\omega}{\omega_{p}}\right)  ^{2}+\dfrac{1}{2}K_{z}^{2}\left(
\frac{\omega}{\omega_{p}}\right)  +K_{r}^{2}=0, \label{c11.2}%
\end{align}
which determine the eigenfrequencies $\omega$ of the various modes. Since the
reduced system has three degrees of freedom, we expect to have three normal
modes. Indeed, when $\omega$ is a solution of the first equation, then
$-\omega$ is a solution of the second equation. We define the mode frequencies
in Eq.(\ref{c11.1}) to be positive (or, in the case of complex $\omega$, to
have positive real part); the negative $\omega$-values are needed to construct
real solutions. Then, the $\Gamma_{+}$-modes describe vibrational motions
turning counter-clockwise coupled to spin precessions turning clockwise, i.e.,
opposite to the natural spin precession, and the $\Gamma_{-}$-modes describe
vibrational motions turning clockwise coupled to spin precessions turning
counter-clockwise, i.e., in the same sense as the natural spin precession.

\begin{figure}[ptb]
\begin{center}
\includegraphics[
height=5.3437in,
width=3.7784in
]{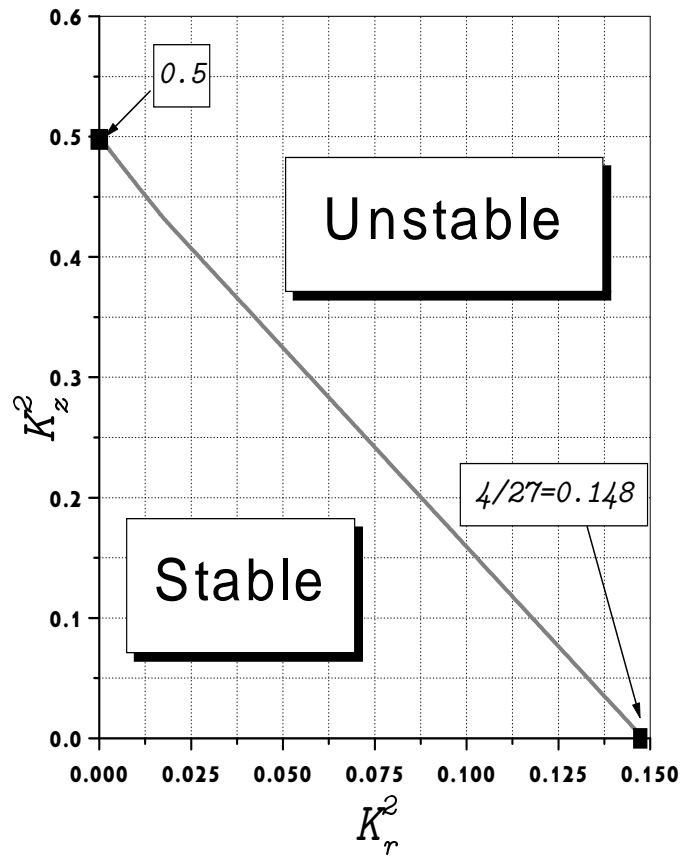}
\caption{Stable region in the $(K_r^2,K_z^2)$-plane, as predicted by the classical analysis.}%
\label{fig1}
\end{center}
\end{figure}

Stability requires that all three solutions of, say, Eq.(\ref{c11.2}) be
\emph{real}. We note that at the edge of the stability region (and when
$K_{r}\neq0$), two out of the three roots of Eq.(\ref{c11.2}) for $\omega$
become identical. In this case, the third order polynomial Eq.(\ref{c11.2})
takes the form $P\left(  \omega\right)  =\left(  \omega-\omega_{1}\right)
^{2}$ $\left(  \omega-\omega_{2}\right)  $, which satisfy $\left.
dP/d\omega\right|  _{\omega=\omega_{1}}=0$. The edge of the stability region
is then found by simultaneously solving the equations $P\left(  \omega\right)
=0$ and $dP/d\omega=0$. The result is given in the form of the parametric
curve in the $(K_{r}^{2},K_{z}^{2})$-plane%
\[
\left\{
\begin{array}
[c]{c}%
K_{r}^{2}\left(  t\right)  =2t^{3}-t^{2}\\
K_{z}^{2}\left(  t\right)  =4t-6t^{2}%
\end{array}
\right\}  \text{ ; with }\dfrac{1}{2}<t<\dfrac{2}{3}%
\]
which is shown in Fig.(\ref{fig1}). Note that by eliminating $t$ from the
second equation and substituting it in the first gives $K_{r}^{2}$
\emph{explicitly} in terms of $K_{z}^{2}$.

\section{Quantum-mechanical analysis.\label{quant}}

\subsection{The Hamiltonian and its diagonalized form.\label{ham}}

In this section we consider the problem of a neutral particle with spin
\emph{one} ($S=\hbar$) in a 3D inhomogeneous magnetic field from a
quantum-mechanical point of view. Unlike the classical analysis, in which the
derivation was valid for any value of the adiabaticity parameters $K_{r}$ and
$K_{z}$, we concentrate here on the behavior of the system when $K_{r}$ and
$K_{z}$ are \emph{small}. Note also that, quantum-mechanically, the magnetic
moment $\mu$ and the spin $S$ of a particle are related by%
\[
\mu=\gamma S,
\]
where $\gamma$ is the gyromagnetic ratio of the particle. Setting $\mu=\gamma
S$ and $S=\hbar$ in Eqs.(\ref{d3}) gives%
\begin{align*}
K_{z}  &  =\dfrac{\omega_{p}}{\omega_{z}}=\sqrt{\dfrac{B^{\prime\prime}\hbar
}{\gamma mB_{0}^{2}}}\\
K_{r}  &  =\dfrac{\omega_{p}}{\omega_{r}}=\sqrt{\dfrac{\left[  (B^{\prime
})^{2}-\frac{1}{2}B_{0}B^{\prime\prime}\right]  \hbar}{\gamma mB_{0}^{3}}}.
\end{align*}

Now, it is convenient to transform to cylindrical coordinates $(r,\phi,z)$ by
setting $x=r\cos\phi$, $y=r\sin\phi$. We denote by $B$ the \emph{amplitude} of
$\mathbf{B}$, by $\theta$ its direction with respect to the $z$-axis and by
$\varphi$ the angle between the projection of $\mathbf{B}$ onto the $\left(
x,y\right)  $-plane and the $x$-axis. Thus, Eq.(\ref{d0}) is rewritten as
\begin{equation}
\mathbf{B}=B\left[  \sin\theta\cos\varphi\mathbf{\hat{x}+}\sin\theta
\sin\varphi\mathbf{\hat{y}}+\cos\theta\mathbf{\hat{z}}\right]  . \label{h1}%
\end{equation}
The approximate expressions for $B$, $\theta$ and $\varphi$ near the origin
are given by
\begin{align}
B\left(  r,\phi,z\right)   &  \simeq B_{0}\left(  1+\dfrac{B_{0}%
B^{\prime\prime}}{2B_{0}^{2}}z^{2}+\left(  \dfrac{(B^{\prime})^{2}-\frac{1}%
{2}B_{0}B^{\prime\prime}}{2B_{0}^{2}}\right)  r^{2}\right)  \text{
}+\mathcal{O}\left(  r^{4},z^{2}r^{2},z^{4}\right)  \text{,}\label{h1.1}\\
\theta\left(  r,\phi,z\right)   &  =\arctan\left(  \dfrac{\sqrt{B_{x}%
^{2}+B_{y}^{2}}}{B_{z}}\right)  \simeq\dfrac{B^{\prime}r}{B_{0}}%
+\mathcal{O}\left(  r^{2},rz,z^{2}\right)  \text{,}\nonumber\\
\varphi\left(  r,\phi,z\right)   &  =\arctan\left(  \dfrac{B_{y}}{B_{x}%
}\right)  \simeq\arctan\left(  -\dfrac{y}{x}\left[  1+\left(  \frac
{B^{\prime\prime}}{B^{\prime}}\right)  z\right]  \right)  \simeq
-\phi+\mathcal{O}\left(  z\sin\left(  2\phi\right)  \right)  .\nonumber
\end{align}
Thus approximately, $B$ and $\theta$ depend only on $r$, whereas $\varphi$
depends only linearly on $\phi$.

The time-independent Schr\"{o}dinger equation for this system is
\begin{equation}
\left[  -\frac{\hbar^{2}}{2m}\nabla^{2}-\mu B\left(  \sin\theta\cos\varphi
\hat{s}_{x}\mathbf{+}\sin\theta\sin\varphi\hat{s}_{y}+\cos\theta\hat{s}%
_{z}\right)  \right]  \Phi(r,\phi,z)=E\Phi(r,\phi,z) \label{h5}%
\end{equation}
where $\hat{s}_{x}$, $\hat{s}_{y}$ and $\hat{s}_{z}$ are the spin one
matrices, given by
\[%
\begin{array}
[c]{ccc}%
\hat{s}_{x}=\dfrac{1}{\sqrt{2}}\left(
\begin{array}
[c]{lll}%
0 & 1 & 0\\
1 & 0 & 1\\
0 & 1 & 0
\end{array}
\right)  & \hat{s}_{y}=\dfrac{1}{\sqrt{2}}\left(
\begin{array}
[c]{lll}%
0 & -i & 0\\
i & 0 & -i\\
0 & i & 0
\end{array}
\right)  & \hat{s}_{z}=\left(
\begin{array}
[c]{lll}%
1 & 0 & 0\\
0 & 0 & 0\\
0 & 0 & -1
\end{array}
\right)  ,
\end{array}
\]
$E$ is the eigenenergy, and $\Phi$ is the three-components spinor
\begin{equation}
\Phi=\left(
\begin{array}
[c]{c}%
\Phi_{+}(r,\phi,z)\\
\Phi_{0}(r,\phi,z)\\
\Phi_{-}(r,\phi,z)
\end{array}
\right)  . \label{h5.1}%
\end{equation}
In matrix form Eq.(\ref{h5}) becomes
\begin{equation}
\left(  H_{K}+H_{M}\right)  \left(
\begin{array}
[c]{c}%
\Phi_{+}(r,\phi,z)\\
\Phi_{0}(r,\phi,z)\\
\Phi_{-}(r,\phi,z)
\end{array}
\right)  =E\left(
\begin{array}
[c]{c}%
\Phi_{+}(r,\phi,z)\\
\Phi_{0}(r,\phi,z)\\
\Phi_{-}(r,\phi,z)
\end{array}
\right)  , \label{h6.0}%
\end{equation}
where $H_{K}$ and $H_{M}$, given by
\begin{align}
H_{K}  &  \equiv-\dfrac{\hbar^{2}}{2m}\nabla^{2}\label{h6.1}\\
H_{M}  &  \equiv-\mu B\left(
\begin{array}
[c]{lll}%
\cos\theta & \dfrac{1}{\sqrt{2}}\sin\theta e^{-i\varphi} & 0\\
\dfrac{1}{\sqrt{2}}\sin\theta e^{i\varphi} & 0 & \dfrac{1}{\sqrt{2}}\sin\theta
e^{-i\varphi}\\
0 & \dfrac{1}{\sqrt{2}}\sin\theta e^{i\varphi} & -\cos\theta
\end{array}
\right)  ,\nonumber
\end{align}
are the kinetic part and the magnetic part of the Hamiltonian $H$, respectively.

In order to diagonalize the magnetic part of the Hamiltonian, we make a local
\emph{passive} transformation of coordinates on the wavefunction such that the
spinor is expressed in a new coordinate system whose $\mathbf{\hat{z}}$ axis
coincides with the direction of the magnetic field at the point $\left(
r,\phi,z\right)  $. We denote by $R\left(  r,\phi,z\right)  $ the required
transformation and set $\Psi=R\Phi$. Thus, $\Psi$ represent \emph{the same}
direction of the spin as before the transformation but using the \emph{new}
coordinate system. The Hamiltonian in this newly defined system is given by
$RHR^{-1}$. We represent the rotation matrix $R$ in terms of the three Euler
angles: First, we perform a rotation through an angle $\varphi$ around the
$\hat{z}$ axis. Second, we make a rotation through an angle $\theta$ around
the \emph{new} position of the $\hat{y}$ axis. At the end of this process the
new $\hat{z}$ axis coincide with the direction of the magnetic field. Now the
value of the last Euler angle, which is a rotation around the new $\hat{z}$
axis, has no effect on this axis. For simplicity we choose this angle to be
$0$. Thus, the representation of the complete transformation for spin-one
particle is given by \cite{rot}
\[
R=\exp\left[  i\theta\hat{s}_{y}\right]  \exp\left[  i\varphi\hat{s}%
_{z}\right]  ,
\]
while its inverse is given by
\[
R^{-1}=\exp\left[  -i\varphi\hat{s}_{z}\right]  \exp\left[  -i\theta\hat
{s}_{y}\right]  .
\]
It is easily verified that the transformation indeed diagonalizes the magnetic
part of the Hamiltonian as%

\[
RH_{M}R^{-1}=-\mu B\hat{s}_{z}.
\]
As for the kinetic part we show at the Appendix that
\begin{equation}
RH_{K}R^{-1}=-\dfrac{\hbar^{2}}{2m}\left[
\begin{array}
[c]{c}%
-i\nabla^{2}\varphi\left(  \cos\theta\hat{s}_{z}-\sin\theta\hat{s}_{x}\right)
-\left|  \nabla\varphi\right|  ^{2}\left(  \cos\theta\hat{s}_{z}-\sin
\theta\hat{s}_{x}\right)  ^{2}\\
-2i\left(  \cos\theta\hat{s}_{z}-\sin\theta\hat{s}_{x}\right)  \nabla
\varphi\cdot\left(  -i\mathbf{\nabla}\theta\hat{s}_{y}+\mathbf{\nabla}\right)
\\
-i\nabla^{2}\theta\hat{s}_{y}-\left|  \mathbf{\nabla}\theta\right|  ^{2}%
\hat{s}_{y}^{2}-2i\hat{s}_{y}\mathbf{\nabla}\theta\cdot\nabla+\nabla^{2}%
\end{array}
\right]  . \label{trans}%
\end{equation}
Since we are interested in the behavior near the origin, we substitute the
approximate expressions Eqs.(\ref{h1.1}) into Eq.(\ref{trans}), replace
$\cos\theta$ by $1$ and $\sin\theta$ by $0$ (since $\theta$ changes very
slowly as compared to the extent over which $\mu B$ changes significantly),
and neglect the terms that are proportional to $\nabla^{2}\theta$ and $\left|
\mathbf{\nabla}\theta\right|  ^{2}$ (being of higher order with respect to
$\mathbf{\nabla}\theta\cdot\nabla$ ). This gives%
\[
RH_{K}R^{-1}\simeq-\dfrac{\hbar^{2}}{2m}\left[  \nabla^{2}-2i\hat{s}_{y}%
\dfrac{B^{\prime}}{B_{0}}\dfrac{\partial}{\partial r}+2i\dfrac{1}{r^{2}}%
\hat{s}_{z}\dfrac{\partial}{\partial\phi}-\dfrac{1}{r^{2}}\hat{s}_{z}%
^{2}\right]  \text{.}%
\]

Thus, the Hamiltonian of the system in the rotated frame may be written
approximately as
\begin{equation}
H\simeq H_{diag}+H_{int}\text{,} \label{h7}%
\end{equation}
where
\begin{align}
H_{diag}  &  =-\dfrac{\hbar^{2}}{2m}\left[  \mathbf{\nabla}^{2}+\dfrac
{2i}{r^{2}}\hat{s}_{z}\dfrac{\partial}{\partial\phi}-\dfrac{1}{r^{2}}\hat
{s}_{z}^{2}\right]  -\mu B\hat{s}_{z}\label{h7.01}\\
H_{int}  &  =i\dfrac{\hbar^{2}B^{\prime}}{mB_{0}}\hat{s}_{y}\dfrac{\partial
}{\partial r}.\nonumber
\end{align}
The first part of the Hamiltonian $H_{diag}$ is diagonal with respect to the
spin degrees of freedom. It contains the kinetic part $\sim\mathbf{\nabla}%
^{2}$, a term whose form is $-\mu B\hat{s}_{z}$ which is identified as the
adiabatic effective potential, and the terms $\sim1/r^{2},ir^{-2}\hat{s}%
_{z}\partial/\partial\phi$ which appear due to the rotation. The second part
of the Hamiltonian $H_{int}$ contains only non-diagonal components. Generally,
$H_{int}$ should contain terms which couple a spin state $M$ to the two
nearest spin states $M\pm1$ and to the two next-to-nearest spin states $M\pm2$
(see the Appendix). In the limit where $K_{z}$ and $K_{r}$ are small, we see
that the coupling of the state with spin projection value $M$ to the states
$M\pm2$ is \emph{negligible} compared to its coupling to the $M\pm1$ states.
We proceed to find the eigenstates of $H_{diag}$.

\subsection{Stationary states of $H_{diag}$.\label{diag}}

Since $H_{diag}$ is diagonal, the three spin states of the wavefunction are
decoupled. We seek a solution for the spin-down ($M=-1$) state%
\begin{equation}
\Psi_{-}=\left(
\begin{array}
[c]{c}%
0\\
0\\
\psi_{-}(r,\phi,z)
\end{array}
\right)  \text{ ; }E=E_{-}, \label{h8.01}%
\end{equation}
and for the $M=0$ state
\begin{equation}
\Psi_{0}=\left(
\begin{array}
[c]{c}%
0\\
\psi_{0}(r,\phi,z)\\
0
\end{array}
\right)  \text{ ; }E=E_{0}. \label{h8.02}%
\end{equation}
We do not consider the \emph{spin-up} ($M=+1$) state, since its coupling to
the trapped spin-down state is negligible, as explained above.

The equation for the non-vanishing component of the spin-down state reads
\begin{equation}
\left\{  -\dfrac{\hbar^{2}}{2m}\left[  \mathbf{\nabla}^{2}-\dfrac{2i}{r^{2}%
}\dfrac{\partial}{\partial\phi}-\dfrac{1}{r^{2}}\right]  +\mu B\right\}
\psi_{-}=E_{-}\psi_{-}, \label{h8.1}%
\end{equation}
whereas the equation for the non-vanishing component of the spin-zero state
is
\begin{equation}
-\dfrac{\hbar^{2}}{2m}\mathbf{\nabla}^{2}\psi_{0}=E_{0}\psi_{0}. \label{h8.2}%
\end{equation}

The solutions of these equations is outlined in the next two subsections.

\subsubsection{Stationary spin-down ($M=-1$) states.\label{down}}

Eq.(\ref{h8.1}) represents a particle in a cylindrically symmetric
\emph{attractive} 3D potential. If the extent of the wave function is small
enough we can expand $B$ in Eq.(\ref{h1.1}) to second order in $r$ and $z$ as
given by Eq.(\ref{h1.1}), and apply the well-known solution of the harmonic
oscillator \cite{sho} in 3D. Under this approximation, Eq.(\ref{h8.1}) becomes%
\begin{equation}
\left\{
\begin{array}
[c]{c}%
-\dfrac{\hbar^{2}}{2m}\left(  \dfrac{\partial^{2}}{\partial z^{2}}%
+\dfrac{\partial^{2}}{\partial r^{2}}+\dfrac{1}{r}\dfrac{\partial}{\partial
r}-\dfrac{1}{r^{2}}\left(  i\dfrac{\partial}{\partial\phi}+1\right)
^{2}\right)  \\
+\left[  \dfrac{m\omega_{z}^{2}}{2}z^{2}+\dfrac{m\omega_{r}^{2}}{2}%
r^{2}\right]
\end{array}
\right\}  \psi_{-}=\left(  E_{-}-\mu B_{0}\right)  \psi_{-}.\label{down0.1}%
\end{equation}
The $z$-coordinate decouples and we assume that it is in the ground-state. We
thus seek a solution whose form is%
\begin{equation}
\psi_{-}(r,z,\phi)=f(r)e^{i\nu\phi}\left(  \dfrac{m\omega_{z}}{\pi\hbar
}\right)  ^{1/4}\exp\left[  -\dfrac{m\omega_{z}z^{2}}{2\hbar}\right]
,\label{down0.2}%
\end{equation}
with $\nu$ an integer. The equation satisfied by $f\left(  r\right)  $ is then%
\begin{equation}
-\dfrac{\hbar^{2}}{2m}\left[  \dfrac{d^{2}f}{dr^{2}}+\dfrac{1}{r}\dfrac
{df}{dr}-\dfrac{f}{r^{2}}\left(  \nu-1\right)  ^{2}\right]  +\dfrac
{m\omega_{r}^{2}r^{2}}{2}f=\left(  E_{-}-\mu B_{0}-\dfrac{1}{2}\hbar\omega
_{z}\right)  f.\label{down0.3}%
\end{equation}
This is an eigenvalue problem for $f$. The smallest eigenvalue for this
equation is obtained by setting%
\[
\nu=1,
\]
for which the eigenfunction $f$ is%
\[
f\left(  r\right)  =De^{i\phi}\exp\left[  -\dfrac{m\omega_{r}}{2\hbar}%
r^{2}\right]  .
\]
Thus, under the harmonic-oscillator approximation, the normalized down-part of
the spin-down state is given by
\begin{equation}
\psi_{-}=\sqrt{\dfrac{m\omega_{r}}{\pi\hbar}}\text{ }\left(  \dfrac
{m\omega_{z}}{\pi\hbar}\right)  ^{1/4}e^{i\phi}\exp\left[  -\dfrac{m\omega
_{r}}{2\hbar}r^{2}\right]  \exp\left[  -\dfrac{m\omega_{z}}{2\hbar}%
z^{2}\right]  .\label{down1}%
\end{equation}
Note that the extent of this wave function over which it changes appreciably
is given by
\begin{equation}
\Delta z\sim\sqrt{K_{z}}\sqrt{\dfrac{B_{0}}{B^{\prime\prime}}}\text{ ; }\Delta
r\sim\sqrt{K_{r}}\sqrt{\dfrac{B_{0}}{B^{\prime\prime}}}\label{down2}%
\end{equation}
whereas the extent over which $\mu B$ changes significantly (see
Eq.(\ref{h1.1})) is
\begin{equation}
\Delta r_{\mu B}\sim\sqrt{\dfrac{B_{0}}{B^{\prime\prime}}}.\label{down3}%
\end{equation}
Thus, the ratio between these two length scales is
\begin{equation}
\dfrac{\Delta z}{\Delta r_{\mu B}}\sim\sqrt{K_{z}}\text{ ; }\dfrac{\Delta
r}{\Delta r_{\mu B}}\sim\sqrt{K_{r}}.\label{down4}%
\end{equation}
We therefore conclude that when $K_{z}$ and $K_{r}$ are small enough, the
harmonic approximation is justified.

The wave function $\psi_{-}$, given by Eq.(\ref{down1}), then represents the
lowest possible \emph{bound} state for this system. This state corresponds to
a \emph{trapped} particle. The energy of this state is
\begin{equation}
E_{-}=\mu B_{0}+\dfrac{1}{2}\hbar\omega_{z}+\hbar\omega_{r}=\mu B_{0}\left(
1+K_{z}+2K_{r}\right)  \simeq\mu B_{0}, \label{down5}%
\end{equation}
while its full spinor representation is
\begin{equation}
\Psi_{-}=\left(
\begin{array}
[c]{c}%
0\\
0\\
\sqrt{\dfrac{m\omega_{r}}{\pi\hbar}}\text{ }\left(  \dfrac{m\omega_{z}}%
{\pi\hbar}\right)  ^{1/4}e^{i\phi}\exp\left[  -\dfrac{m\omega_{r}}{2\hbar
}r^{2}\right]  \exp\left[  -\dfrac{m\omega_{z}}{2\hbar}z^{2}\right]
\end{array}
\right)  . \label{down6}%
\end{equation}

\subsubsection{Stationary ($M=0$) states.\label{up}}

Eq.(\ref{h8.2}) describes a free particle. It corresponds to an unbounded
state representing an \emph{untrapped} particle. In this case there is a
continuum of states, each with its own energy. As we are interested in
non-radiative decay, we focus on finding a solution with an energy which is
\emph{equal} to the energy found for the trapped state, that is
\begin{equation}
E_{0}=E_{-}\simeq\mu B_{0}. \label{up0}%
\end{equation}
We seek a solution in the form%
\[
\psi_{0}(r,\phi)=g(r)\exp\left[  ik_{z}z+i\beta\phi\right]  ,
\]
where $\beta$ is an integer. Substituting this, together with Eq.(\ref{up0})
into Eq.(\ref{h8.2}) gives%
\[
\left[  \dfrac{d^{2}}{dr^{2}}+\dfrac{1}{r}\dfrac{d}{dr}+\left(  k_{r}%
^{2}-\dfrac{\beta^{2}}{r^{2}}\right)  \right]  g=0,
\]
where%
\[
k_{r}^{2}+k_{z}^{2}=\dfrac{2\mu mB_{0}}{\hbar^{2}}\text{.}%
\]
The non-singular solution for $g$ is%
\[
g\left(  r\right)  =J_{\beta}\left(  k_{r}r\right)  ,
\]
where $J_{\beta}\left(  x\right)  $ is the Bessel function of the first kind
of order $\beta$. For what follows, it is convenient to introduce an angle
$\gamma$ such that%
\begin{align*}
k_{r}  &  =k_{0}\sin\gamma\\
k_{z}  &  =k_{0}\cos\gamma.\\
k_{0}  &  \equiv\sqrt{\dfrac{2\mu mB_{0}}{\hbar^{2}}}%
\end{align*}
with $0<\gamma<\pi$.

We note that $H_{int}$ does not operate on the $\phi$ coordinate. Hence, in
order to have a non-vanishing matrix element between the zero-state and the
down-state, they must have the \emph{same} $\phi$-dependence. Thus, $\beta
=\nu=1$, and as a result, the state with angle $\gamma$ is given by
\begin{equation}
\psi_{0}^{\gamma}(r,\phi,z)=C_{\gamma}J_{1}\left(  k_{0}r\sin\gamma\right)
\exp\left[  i\left(  \phi+k_{0}z\cos\gamma\right)  \right]  \text{.}
\label{up0.2}%
\end{equation}
with%
\begin{equation}
\Psi_{0}^{\gamma}=\left(
\begin{array}
[c]{c}%
0\\
C_{\gamma}J_{1}\left(  k_{0}r\sin\gamma\right)  \exp\left[  i\left(
\phi+k_{0}z\cos\gamma\right)  \right] \\
0
\end{array}
\right)  ,\text{ } \label{up0.3}%
\end{equation}
where $C_{\gamma}$ is the normalization constant which is chosen to be real,
and depends on $\gamma$. To evaluate $C_{\gamma}$ we temporarily introduce
boundary conditions under which the wavefunction $\Psi_{0}^{\gamma}$ vanishes
at $r=R$, and satisfies periodic boundary conditions along $z$ with period
$Z$. Thus, normalization of $\Psi_{0}^{\gamma}$ gives
\[
\int_{-Z/2}^{Z/2}dz\int_{0}^{2\pi}d\phi\int_{0}^{R}rdr\left|  \Psi_{0}%
^{\gamma}\right|  ^{2}=C_{\gamma}^{2}Z2\pi\frac{1}{2}R^{2}[J_{2}(k_{0}%
R\sin\gamma)]^{2}=1,
\]
such that%
\[
C_{\gamma}=\dfrac{1}{\sqrt{Z\pi}R\left|  J_{2}(k_{0}R\sin\gamma)\right|  },
\]
where we have used \cite{grad}
\[
\int_{0}^{R}[J_{1}(kr)]^{2}rdr=\frac{1}{2}R^{2}[J_{2}(kR)]^{2}.
\]
In the asymptotic region $kR\gtrsim1$, the function $J_{2}(kR)$ takes the
values $\pm\sqrt{2/(\pi kR)}$ at the zeros of $J_{1}(kR)$. Thus,%
\[
C_{\gamma}=\dfrac{1}{\sqrt{Z\pi}R\left|  J_{2}(k_{0}R\sin\gamma)\right|
}\simeq\dfrac{\sqrt{k_{0}\sin\gamma}}{\sqrt{2ZR}}%
\]
and hence%
\begin{equation}
C_{\gamma}^{2}\simeq\dfrac{k_{0}\sin\gamma}{2ZR}. \label{cg2}%
\end{equation}

\subsection{The transition rate.\label{time}}

We calculate the transition rate from the bound state given by Eq.(\ref{down6}%
) to the unbounded state Eq.(\ref{up0.3}), according to Fermi's golden rule
\cite{fermi}. Thus, the infinitesimal decay time from the trapped state to the
untrapped state defined by $\gamma$ is given by
\begin{equation}
d\left(  \dfrac{1}{T_{esc}^{\gamma}}\right)  =\dfrac{2\pi}{\hbar}\left|
H_{i}^{\gamma}\right|  ^{2}\rho_{\gamma}(E_{0})d\gamma,\label{t1}%
\end{equation}
where $\rho_{\gamma}(E)d\gamma$ is the density $dN/dE$ of states $\psi
_{0}^{\gamma}$ with an angle between $\gamma$ and $\gamma+d\gamma$ and energy
between $E_{0}$ and $E_{0}+dE$, and $H_{i}^{\gamma}$ is the matrix element of
$H_{int}$ between the bound state and the unbounded state defined by $\gamma$
and $E_{0}$. To find $\rho_{\gamma}(E_{0})$ we note that the final state is
defined by the two quantized $k$-vectors $k_{r}=k_{0}\sin\gamma$ and
$k_{z}=k_{0}\cos\gamma$. The possible $k_{z}$ values are equally-spaced with
lattice constant $dk_{z}=2\pi/Z$. Since the Bessel function $J_{1}(k_{r}r)$ is
very close to its asymptotic behavior at large arguments
\[
J_{1}(k_{r}R)\simeq\sqrt{\frac{2}{\pi k_{r}R}}\,\cos\left(  k_{r}R-\frac{3\pi
}{4}\right)  ,
\]
the $k_{r}$ are also very much equally-spaced (even when $k_{r}$ is small, it
is still a good approximation) with lattice constant $dk_{r}\simeq\pi/R$.
Thus, in the $\left(  k_{r},k_{z}\right)  $-space, the allowed $k$ vectors
form a regular lattice, and the number of states $dN$ in the volume element
$k_{0}dk_{0}d\gamma$ is given by%
\[
dN=\dfrac{k_{0}d\gamma dk_{0}}{\left(  \dfrac{\pi}{R}\right)  \left(
\dfrac{2\pi}{Z}\right)  }.
\]
With $dE=\hbar^{2}k_{0}dk_{0}/m$, this gives the density of states%
\[
\rho_{\gamma}(E)d\gamma=\dfrac{dN}{dE}\simeq\dfrac{mZR}{2\pi^{2}\hbar^{2}%
}d\gamma.
\]
This, together with Eq.(\ref{cg2}) yields%
\begin{equation}
C_{\gamma}^{2}\rho_{\gamma}(E)\simeq\dfrac{mk_{0}\sin\gamma}{4\pi^{2}\hbar
^{2}}.\label{c2rho}%
\end{equation}
Evaluation of $H_{i}^{\gamma}$ gives%
\begin{align*}
H_{i}^{\gamma} &  =\int_{-\infty}^{\infty}dz%
{\displaystyle\int\limits_{0}^{\infty}}
rdr%
{\displaystyle\int\limits_{0}^{2\pi}}
d\phi\Psi_{0}^{\dagger\gamma}H_{int}\Psi_{-}\\
&  =2\pi\sqrt{\dfrac{m\omega_{r}}{\pi\hbar}}\text{ }\left(  \dfrac{m\omega
_{z}}{\pi\hbar}\right)  ^{1/4}\left[  i\dfrac{\hbar^{2}B^{\prime}}{mB_{0}%
}\dfrac{\left(  -i\right)  }{\sqrt{2}}\right]  \left(  -\dfrac{m\omega_{r}%
}{\hbar}\right)  C_{\gamma}\\
\times &
{\displaystyle\int\limits_{0}^{\infty}}
drr^{2}J_{1}\left(  k_{0}r\sin\gamma\right)  \exp\left[  -\dfrac{m\omega_{r}%
}{2\hbar}r^{2}\right]  \int_{-\infty}^{\infty}dz\exp\left[  -\dfrac
{m\omega_{z}}{2\hbar}z^{2}-ik_{0}z\cos\gamma\right]  .
\end{align*}
and hence%
\begin{align}
\left|  H_{i}^{\gamma}\right|  ^{2} &  =4\pi^{2}\dfrac{m\omega_{r}}{\pi\hbar
}\sqrt{\dfrac{m\omega_{z}}{\pi\hbar}}\left(  \dfrac{\hbar^{2}B^{\prime}}%
{\sqrt{2}mB_{0}}\right)  ^{2}\left(  \dfrac{m\omega_{r}}{\hbar}\right)
^{2}C_{\gamma}^{2}\dfrac{2\pi\hbar}{m\omega_{z}}\left(  \dfrac{k_{0}\hbar
^{2}\sin\gamma}{m^{2}\omega_{r}^{2}}\right)  ^{2}\label{hi2}\\
&  \times\exp\left[  -\frac{\hbar k_{0}^{2}\cos^{2}\gamma}{m\omega_{z}}%
-\dfrac{k_{0}^{2}\hbar\sin^{2}\gamma}{m\omega_{r}}\right]  .\nonumber
\end{align}
Substituting Eqs.(\ref{hi2}) and (\ref{c2rho}) into Eq.(\ref{t1}) and
integrating over $\gamma$ from $0$ to $\pi$ gives%

\begin{equation}
\dfrac{1}{T_{esc}}=2\sqrt{2\pi}\frac{\left(  2\omega_{r}^{2}+\omega_{z}%
^{2}\right)  \sqrt{\omega_{p}}}{\omega_{r}\sqrt{\omega_{z}}}I\left(
\dfrac{2\omega_{p}}{\omega_{r}},\dfrac{2\omega_{p}}{\omega_{z}}\right)
\label{1/TT}%
\end{equation}
with%
\begin{equation}
I\left(  a,b\right)  \equiv\int_{0}^{\pi}d\gamma\sin^{3}\gamma\exp\left[
-a\sin^{2}\gamma-b\cos^{2}\gamma\right]  ,\label{I}%
\end{equation}
where we have substituted our previous definitions for $\omega_{p}$,
$\omega_{r}$ and $\omega_{z}$. The integral $I\left(  a,b\right)  $ may be
expressed in terms of the simpler integral%
\begin{align}
I_{0}\left(  a,b\right)   &  \equiv\int_{0}^{\pi}d\gamma\sin\gamma\exp\left[
-a\sin^{2}\gamma-b\cos^{2}\gamma\right]  \label{I0}\\
&  =2\exp\left(  -a\right)  \int_{0}^{1}\exp\left[  \left(  a-b\right)
t^{2}\right]  dt\nonumber
\end{align}
by%
\begin{equation}
I\left(  a,b\right)  =-\dfrac{\partial I_{0}\left(  a,b\right)  }{\partial
a}.\label{I2}%
\end{equation}
In the isotropic case where $\omega_{r}=\omega_{z}\equiv\omega_{i}$, the
integral in Eq.(\ref{1/TT}) can be evaluated analytically with the result that%
\[
\dfrac{1}{T_{esc}}=8\sqrt{2\pi}\sqrt{\omega_{p}\omega_{i}}\exp\left[
-\frac{2\omega_{p}}{\omega_{i}}\right]  .
\]
In the extreme cases $\omega_{r}\gg\omega_{z}$ and $\omega_{z}\gg\omega_{r}$
we obtain from the asymptotic behavior of the error function of real and
imaginary argument \cite{abram}%
\begin{equation}
I_{0}\left(  a,b\right)  \simeq\left\{
\begin{array}
[c]{cc}%
\sqrt{\dfrac{\pi}{b}}\exp\left(  -a\right)   & \text{; }b\gg a\\
\dfrac{1}{a}\exp\left(  -b\right)   & \text{; }b\ll a
\end{array}
\right.  ,\label{i0app}%
\end{equation}
and hence%
\begin{equation}
I\left(  a,b\right)  =-\dfrac{\partial I_{0}\left(  a,b\right)  }{\partial
a}\simeq\left\{
\begin{array}
[c]{cc}%
\sqrt{\dfrac{\pi}{b}}\exp\left(  -a\right)   & \text{; }b\gg a\\
\dfrac{1}{a^{2}}\exp\left(  -b\right)   & \text{; }b\ll a
\end{array}
\right.  .\label{i2}%
\end{equation}
Substituting Eqs.(\ref{i0app}) and (\ref{i2}) into Eq.(\ref{1/TT}) gives%
\begin{equation}
\dfrac{1}{T_{esc}}\simeq\left\{
\begin{array}
[c]{c}%
4\pi\omega_{r}\exp\left[  -\dfrac{2\omega_{p}}{\omega_{r}}\right]  \text{; for
}\omega_{p}\gg\omega_{r}\gg\omega_{z}\\
8\sqrt{2\pi}\sqrt{\omega_{p}\omega_{i}}\exp\left[  -\dfrac{2\omega_{p}}%
{\omega_{i}}\right]  \text{ ; for }\omega_{p}\gg\omega_{r}=\omega_{z}%
\equiv\omega_{i}\\
\sqrt{\dfrac{\pi}{2}}\omega_{r}\left(  \dfrac{\omega_{z}}{\omega_{p}}\right)
^{3/2}\exp\left[  -\dfrac{2\omega_{p}}{\omega_{z}}\right]  \text{; for }%
\omega_{p}\gg\omega_{z}\gg\omega_{r}%
\end{array}
\right.  ,\label{final}%
\end{equation}
with the conclusion that the transition rate is dominated by the
\emph{largest} of the two frequencies $\omega_{z}$ and $\omega_{r}$.

\section{Discussion.\label{dis}}

The problem we have studied has three important time scales: The shortest time
scale is $T_{prec}$, which is the time required for \emph{one} precession of
the spin around the axis of the local magnetic field. The intermediate time
scale is given by $T_{r,z}=T_{prec}/K_{r,z}$, which are the times required to
complete one cycle of the center of mass around the center of the trap in the
lateral and axial directions, respectively. These two time scales appear both
in the classical and the quantum-mechanical analysis. The longest time scale
(provided that $K_{r}$ and $K_{z}$ are small) $T_{esc}$, which is not present
in the classical problem, is the time it takes for the particle to escape from
the trap.

Whereas the classical analysis yields an upper bound for $K_{z}$ and $K_{r}$
for trapping to occur, no such sharp bound exists in the quantum-mechanical
analysis. Nevertheless it is interesting to compare the classical bound with
the values of $K_{z}$ and $K_{r}$ for which the exponent in the expression for
the quantum-mechanical lifetime becomes equal to $1$: According to
Fig.(\ref{fig1}), we find that $K_{z},_{\max}=1/\sqrt{2}=0.707$ when $K_{r}%
=0$, and $K_{r,\max}=\sqrt{4/27}=0.385$ when $K_{z}=0$. From Eq.(\ref{final})
on the other hand, we conclude that $K_{z},_{cr}=0.5$ when $K_{r}=0$, and
$K_{r,cr}=0.5$ when $K_{z}=0$. Thus, the quantum-mechanical condition for
trapping to occur is roughly the same as the classical condition. These
results however, should be taken with caution since our quantum-mechanical
analysis is valid only for small values of $K_{r}$ and $K_{z}$.

Though our derivation was for the case of a spin-one particle, it is clear
that it can be extended to particles with higher spin, and also to
\emph{half-integer} spin particles. In view of the results obtained by our
recent study of spin half particles in 1D field \cite{life1d} and 2D field
\cite{life2d}, we believe that the expression for the lifetime in these cases
is similar to the result which is obtained in the present paper.

As an example, we apply our results to the case of a spin $1$ atom that is
trapped in a field with $B_{0}=100$ Oe and $B_{0}/B^{\prime}\sim\sqrt
{B_{0}/B^{\prime\prime}}\sim10$cm. These parameters correspond to typical
traps used in Bose-Einstein condensation experiments \cite{bec,bec2,bec3,bec4}%
. The results, being correct to within an order of magnitude, are outlined in
Table \ref{tab1}. We note that in both cases the values of $K_{z}$ and $K_{r}$
are much smaller than $1$. Also, the calculated lifetime of the particle in
the trap is extremely large, suggesting that the particle is tightly trapped
in this field.

In this study we have been interested in the \emph{ground-state }trapped
state. In the case of a particle with spin $1/2$ or spin $1$, this is the only
\emph{one} trapped spin state. However, when particles with higher spin are
considered there are more than one trapped states. A natural question in
connection with these is what is the lifetime of these trapped states. Another
interesting issue is the lifetime of an \emph{excited} state in a given
trapped spin state. Our preliminary results show that some of these excited
states may have a \emph{short} lifetime, being \emph{algebraically} dependent
on $\omega_{p}/\omega_{r}$ and $\omega_{p}/\omega_{z}$ rather than
\emph{exponentially} dependent. This question is still under study.

\pagebreak 

\bigskip\appendix

\section{Transformation of $\nabla^{2}$.}

\bigskip The transformation of $\nabla^{2}$ is given by
\begin{equation}
R\nabla^{2}R^{-1}=\exp\left[  i\theta\hat{s}_{y}\right]  Q\exp\left[
-i\theta\hat{s}_{y}\right]  , \label{a1}%
\end{equation}
where%
\begin{equation}
Q\equiv\exp\left[  i\varphi\hat{s}_{z}\right]  \nabla^{2}\exp\left[
-i\varphi\hat{s}_{z}\right]  . \label{a2}%
\end{equation}
Evaluating $Q$ first gives%
\begin{align*}
\nabla^{2}\left(  \exp\left[  -i\varphi\hat{s}_{z}\right]  A\right)   &
=\mathbf{\nabla}\cdot\mathbf{\nabla}\left(  \exp\left[  -i\varphi\hat{s}%
_{z}\right]  A\right) \\
&  =\mathbf{\nabla}\cdot\left[  \mathbf{\nabla}\left(  \exp\left[
-i\varphi\hat{s}_{z}\right]  \right)  A+\exp\left[  -i\varphi\hat{s}%
_{z}\right]  \mathbf{\nabla}A\right] \\
&  =\mathbf{\nabla}\cdot\left[  -i\exp\left[  -i\varphi\hat{s}_{z}\right]
\mathbf{\nabla}\varphi\hat{s}_{z}A+\exp\left[  -i\varphi\hat{s}_{z}\right]
\mathbf{\nabla}A\right] \\
&  =\mathbf{\nabla}\cdot\left[  -i\exp\left[  -i\varphi\hat{s}_{z}\right]
\mathbf{\nabla}\varphi\hat{s}_{z}A+\exp\left[  -i\varphi\hat{s}_{z}\right]
\mathbf{\nabla}A\right]
\end{align*}
but%
\begin{align*}
&  \mathbf{\nabla}\cdot\left[  -i\exp\left[  -i\varphi\hat{s}_{z}\right]
\mathbf{\nabla}\varphi\hat{s}_{z}A\right] \\
&  =\mathbf{\nabla}\varphi\cdot\mathbf{\nabla}\left[  -i\exp\left[
-i\varphi\hat{s}_{z}\right]  \hat{s}_{z}A\right]  -i\exp\left[  -i\varphi
\hat{s}_{z}\right]  \mathbf{\nabla}^{2}\varphi\hat{s}_{z}A\\
&  =\mathbf{\nabla}\varphi\cdot\left[  -i\exp\left[  -i\varphi\hat{s}%
_{z}\right]  \hat{s}_{z}\mathbf{\nabla}A-\exp\left[  -i\varphi\hat{s}%
_{z}\right]  \mathbf{\nabla}\varphi\hat{s}_{z}^{2}A\right]  -i\exp\left[
-i\varphi\hat{s}_{z}\right]  \mathbf{\nabla}^{2}\varphi\hat{s}_{z}A
\end{align*}
and%
\begin{align*}
&  \mathbf{\nabla}\cdot\left[  \exp\left[  -i\varphi\hat{s}_{z}\right]
\mathbf{\nabla}A\right] \\
&  =-i\mathbf{\nabla}\varphi\exp\left[  -i\varphi\hat{s}_{z}\right]  \hat
{s}_{z}\cdot\mathbf{\nabla}A+\exp\left[  -i\varphi\hat{s}_{z}\right]
\mathbf{\nabla}^{2}A
\end{align*}
hence%
\begin{align*}
&  \nabla^{2}\left(  \exp\left[  -i\varphi\hat{s}_{z}\right]  A\right) \\
&  =\mathbf{\nabla}\varphi\cdot\left[  -i\exp\left[  -i\varphi\hat{s}%
_{z}\right]  \hat{s}_{z}\mathbf{\nabla}A-\exp\left[  -i\varphi\hat{s}%
_{z}\right]  \mathbf{\nabla}\varphi\hat{s}_{z}^{2}A\right]  -i\exp\left[
-i\varphi\hat{s}_{z}\right]  \mathbf{\nabla}^{2}\varphi\hat{s}_{z}A\\
&  -i\mathbf{\nabla}\varphi\exp\left[  -i\varphi\hat{s}_{z}\right]  \hat
{s}_{z}\cdot\mathbf{\nabla}A+\exp\left[  -i\varphi\hat{s}_{z}\right]
\mathbf{\nabla}^{2}A\\
&  =-2i\mathbf{\nabla}\varphi\exp\left[  -i\varphi\hat{s}_{z}\right]  \hat
{s}_{z}\cdot\mathbf{\nabla}A-\exp\left[  -i\varphi\hat{s}_{z}\right]  \left|
\mathbf{\nabla}\varphi\right|  ^{2}\hat{s}_{z}^{2}A\\
&  -i\exp\left[  -i\varphi\hat{s}_{z}\right]  \mathbf{\nabla}^{2}\varphi
\hat{s}_{z}A+\exp\left[  -i\varphi\hat{s}_{z}\right]  \mathbf{\nabla}^{2}A
\end{align*}
thus%
\begin{align*}
&  \exp\left[  i\varphi\hat{s}_{z}\right]  \nabla^{2}\left(  \exp\left[
-i\varphi\hat{s}_{z}\right]  A\right) \\
&  =-2i\hat{s}_{z}\mathbf{\nabla}\varphi\cdot\mathbf{\nabla}A-\left|
\mathbf{\nabla}\varphi\right|  ^{2}\hat{s}_{z}^{2}A-i\mathbf{\nabla}%
^{2}\varphi\hat{s}_{z}A+\mathbf{\nabla}^{2}A
\end{align*}
or in an operatorial form%
\begin{equation}
Q=\exp\left[  i\varphi\hat{s}_{z}\right]  \nabla^{2}\exp\left[  -i\varphi
\hat{s}_{z}\right]  =-2i\hat{s}_{z}\mathbf{\nabla}\varphi\cdot\mathbf{\nabla
}-\left|  \mathbf{\nabla}\varphi\right|  ^{2}\hat{s}_{z}^{2}-i\mathbf{\nabla
}^{2}\varphi\hat{s}_{z}+\mathbf{\nabla}^{2}. \label{a3}%
\end{equation}
Substituting in Eq.(\ref{a1}) each of the four terms in Eq.(\ref{a3}) we find%
\begin{equation}
\exp\left[  i\theta\hat{s}_{y}\right]  \nabla^{2}\exp\left[  -i\theta\hat
{s}_{y}\right]  =-i\nabla^{2}\theta\hat{s}_{y}-\left|  \mathbf{\nabla}%
\theta\right|  ^{2}\hat{s}_{y}^{2}-2i\hat{s}_{y}\mathbf{\nabla}\theta
\cdot\nabla+\nabla^{2}\text{,} \label{a3.1}%
\end{equation}%
\begin{equation}
\exp\left[  i\theta\hat{s}_{y}\right]  \left(  -i\mathbf{\nabla}^{2}%
\varphi\right)  \hat{s}_{z}\exp\left[  -i\theta\hat{s}_{y}\right]
=-i\mathbf{\nabla}^{2}\varphi\left(  \cos\theta\hat{s}_{z}-\sin\theta\hat
{s}_{x}\right)  \text{,} \label{a3.2}%
\end{equation}%
\begin{align}
\exp\left[  i\theta\hat{s}_{y}\right]  \hat{s}_{z}^{2}\exp\left[  -i\theta
\hat{s}_{y}\right]   &  =\left(  \cos\theta\hat{s}_{z}-\sin\theta\hat{s}%
_{x}\right)  ^{2}\label{a3.3}\\
&  =\cos^{2}\theta\hat{s}_{z}^{2}-\sin\theta\cos\theta\left(  \hat{s}_{z}%
\hat{s}_{x}+\hat{s}_{x}\hat{s}_{z}\right)  +\sin^{2}\theta\hat{s}_{x}%
^{2}\text{,}\nonumber
\end{align}%
\begin{align}
\exp\left[  i\theta\hat{s}_{y}\right]  \hat{s}_{z}\mathbf{\nabla}\exp\left[
-i\theta\hat{s}_{y}\right]   &  =\exp\left[  i\theta\hat{s}_{y}\right]
\hat{s}_{z}\exp\left[  -i\theta\hat{s}_{y}\right]  \left(  -i\mathbf{\nabla
}\theta\hat{s}_{y}+\mathbf{\nabla}\right) \label{a3.4}\\
&  =\left(  \cos\theta\hat{s}_{z}-\sin\theta\hat{s}_{x}\right)  \left(
-i\mathbf{\nabla}\theta\hat{s}_{y}+\mathbf{\nabla}\right) \nonumber\\
&  =-i\cos\theta\mathbf{\nabla}\theta\hat{s}_{z}\hat{s}_{y}+i\sin
\theta\mathbf{\nabla}\theta\hat{s}_{x}\hat{s}_{y}+\cos\theta\hat{s}%
_{z}\mathbf{\nabla}-\sin\theta\hat{s}_{x}\mathbf{\nabla,}\nonumber
\end{align}
Substituting Eqs.(\ref{a3.1}) to (\ref{a3.4}) into Eq.(\ref{a1}) gives%
\[
R\nabla^{2}R^{-1}=\left[
\begin{array}
[c]{c}%
-i\nabla^{2}\varphi\left(  \cos\theta\hat{s}_{z}-\sin\theta\hat{s}_{x}\right)
-\left|  \nabla\varphi\right|  ^{2}\left(  \cos\theta\hat{s}_{z}-\sin
\theta\hat{s}_{x}\right)  ^{2}\\
-2i\left(  \cos\theta\hat{s}_{z}-\sin\theta\hat{s}_{x}\right)  \nabla
\varphi\cdot\left(  -i\mathbf{\nabla}\theta\hat{s}_{y}+\mathbf{\nabla}\right)
\\
-i\nabla^{2}\theta\hat{s}_{y}-\left|  \mathbf{\nabla}\theta\right|  ^{2}%
\hat{s}_{y}^{2}-2i\hat{s}_{y}\mathbf{\nabla}\theta\cdot\nabla+\nabla^{2}%
\end{array}
\right]  .
\]
Note that the transformed $\nabla^{2}$ is composed of terms containing
$\hat{s}_{i}^{n}$ with $n=0,1$ or $2$. This is a consequence of the fact that
the original operator $\nabla^{2}$ is a second order differential operator.
Thus, a spin state $\Psi_{M}$ for which $\hat{s}_{z}\Psi_{M}=M\Psi_{M}$ is
coupled, in first order, only to the states $\Psi_{M\pm1}$ and $\Psi_{M\pm2}$.

\pagebreak 

\begin{center}%
\begin{table}[tbp] \centering
\caption{Typical time scales for a spin $1$ atom trapped with a field $B_{0}%
=100$ Oe and $B_{0}/B^{\prime}\sim\sqrt{B_{0}/B^{\prime\prime}}\sim10$cm.}%
\begin{tabular}
[c]{|l|l|}\hline
& Spin $1$ atom\\\hline
$m$ gr & $\sim10^{-22}$\\
$\mu$ emu & $\sim10^{-20}$\\
$K_{z}$, $K_{r}$ & $\sim10^{-8}$\\
$\omega_{p}^{-1}$ sec & $\sim10^{-9}$\\
$\omega_{r}^{-1}$, $\omega_{z}^{-1}$ sec & $\sim10^{-1}$\\
$T_{esc}$ sec & $\sim10^{\left(  10^{8}\right)  }$\\\hline
\end{tabular}%
\label{tab1}
\end{table}%
\end{center}

\newpage

\end{document}